*Article*

# Photo-thermal tuning of graphene oxide coated integrated optical waveguides

**David J. Moss**

Optical Sciences Center, Swinburne University of Technology, Hawthorn, VIC 3122, Australia;

**Abstract**—We experimentally investigate power-sensitive photo-thermal tuning (PTT) of two-dimensional (2D) graphene oxide (GO) films coated on integrated optical waveguides. We measure the light power thresholds for reversible and permanent GO reduction in silicon nitride (SiN) waveguides integrated with 1 and 2 layers of GO. Raman spectra at different positions of a hybrid waveguide with permanently reduced GO are characterized, verifying the inhomogeneous GO reduction along the direction of light propagation through the waveguide. The differences between the PTT induced by a continuous-wave laser and a pulsed laser are also compared, confirming that the PTT mainly depend on the average input power. These results reveal interesting features for 2D GO films coated on integrated optical waveguides, which are of fundamental importance for the control and engineering of GO's properties in hybrid integrated photonic devices.

Keywords: 2D materials, integrated optics, photo-thermal changes, graphene oxide.

## 1. Introduction

Graphene oxide (GO) is a two-dimensional (2D) material that has attracted significant interest because of its many attractive properties such as broadband photoluminescence [1, 2], high Kerr nonlinearity [3, 4], strong material anisotropy [5, 6], broadband light absorption [7, 8], and tunable material properties in wide ranges [9, 10]. In addition, with its facile fabrication processes, GO has a strong capability for large-scale manufacturable on-chip integration [6, 11, 12].

The incorporation of GO into integrated photonic devices has led to the birth of GO integrated photonics, which has become a very active and fast-growing field [13]. Integrated photonic devices incorporating GO films have been demonstrated for a range of applications, such as light absorbers [7, 8, 14], optical lenses and imaging devices [10, 11, 15], polarization-selective devices [6, 16], sensors [17, 18], and nonlinear optical devices [19-23].

Since GO can be converted to a reduced form with graphene-like properties under strong light irradiation or high temperature [24, 25], it has long been used as a precursor for the preparation of graphene [13, 26, 27]. Given the difference between the material properties of GO and reduced GO (rGO) [3, 11, 28], investigating the mechanisms and conditions for GO reduction in hybrid integrated photonic devices is of fundamental importance for applying this functional 2D material to integrated photonic devices [13, 29].

Previously, we observed power-sensitive photo-thermal changes in GO films coated on integrated photonic waveguides [19] and ring resonators [30] in nonlinear four-wave mixing experiments. In this paper, we provide a more detailed characterization for such changes arising from a range of effects such as photo-thermal reduction, thermal dissipation, and self-heating in GO layers. We experimentally investigate photo-thermal tuning (PTT) of 2D GO films coated on integrated optical waveguides. We measure the light power thresholds for reversible and permanent GO reduction in silicon nitride (SiN) waveguides integrated with 1 and 2 layers of GO. We identify three reduction stages according to the existence of reversible versus permanent reduction. Raman spectra at different positions of a hybrid waveguide with permanently reduced GO film are also characterized, showing the inhomogeneous nature of GO reduction the direction of light propagation through the waveguide. Finally, we compare the PTT induced by a continuous-wave (CW) laser and a pulsed laser with the same average power, and observe negligible difference between them. This confirms that the PTT mainly depend on the average power rather than the peak power of input light. These results reveal interesting features for the reduction of GO induced by the photo-thermal changes, which are useful for controlling and engineering GO's material properties in hybrid integrated photonic devices.

## 2. Device design and fabrication

**Figure 1a** show a schematic illustration of a GO-coated SiN waveguide with a monolayer GO film. The bare SiN waveguide has a cross-section of 1.70 μm × 0.72 μm. A complementary metal-oxide-semiconductor (CMOS) compatible crack-free method [31-33] was utilized to fabricate the uncoated SiN waveguides. First, a two-step deposition of SiN films was achieved via low-pressure chemical vapor deposition (LPCVD) for strain management and crack prevention. Next, 248-nm deep ultraviolet lithography and $CF_4/CH_2F_2/O_2$ fluorine-based dry etching were employed for patterning the SiN waveguides. After that, a silica upper cladding was deposited using high-density plasma-enhanced chemical vapor deposition (HDP-PECVD), followed by opening a window on it down to the top surface of the SiN waveguides via lithography and dry etching processes. The window was located near the waveguide input to enable a relatively high optical power injected into the GO coated segment [34]. The length of the opened window was $L_w$ = 1.4 mm. Finally, 2D layered GO films were coated onto the SiN waveguide by using a solution-based method that enabled transfer-free and layer-by-layer film coating, as reported previously [6, 11, 12, 35]. Compared to the cumbersome film transfer processes employed for the on-chip integration of other 2D materials such as graphene and TMDCs [36-38], our GO coating method is highly scalable, enabling the precise control of the GO layer number (i.e., film thickness), the ability to coat large-area films, and good film attachment onto integrated chips [11, 13]. The total length of the SiN waveguide was $L$ = 2 cm. A micrograph of the fabricated device corresponding to **Figure 1a** is shown in **Figure 1b**. The GO film coated on the chip surface exhibited good morphology, high transmittance, and high uniformity. **Figure 1c** shows the schematic cross section of the hybrid waveguide in **Figure 1a**, the corresponding TE mode profile is shown in **Figure 1d**. The light-matter interaction between the waveguide evanescent field and the GO film can induce power-sensitive photo-thermal changes in the GO film, which has been observed previously [19, 29, 30]. In this paper, we only investigate PTT induced by the TE polarized light. This is because the TE polarization supports the in-plane interaction between the waveguide evanescent field and the 2D GO film, which is much stronger compared to the out-of-plane interaction given the significant anisotropy of 2D GO films [6], thus allowing for the excitation of higher levels of photo thermal changes.

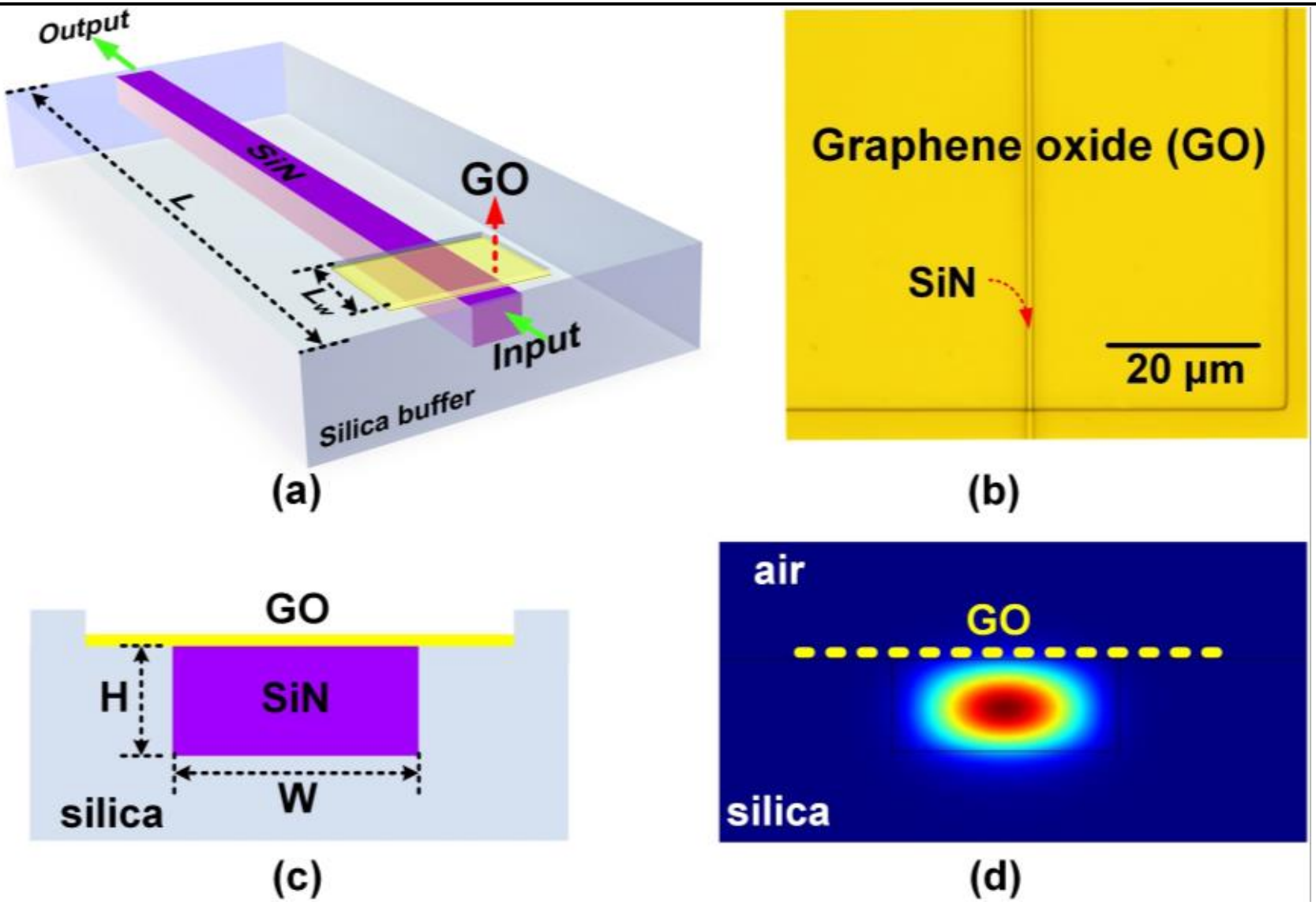

**Figure 1.** (a) Schematic illustration of a SiN waveguide coated with 1 layer of GO. (b) A micrograph showing the area around the opened window of the fabricated device corresponding to (a). (c) Schematic illustration of the cross-section of the hybrid waveguide. (d) TE mode profile corresponding to (c).

## 3. Experimental results

**Figure 2** shows the experimental setup for characterizing PTT of GO films coated on integrated optical waveguides. Two CW light sources having different powers were employed. The low power one was used to measure the loss at low powers without obvious photo thermal changes in the GO films, whereas the high-power one was achieved by amplification with an erbium-doped fiber amplifier (EDFA) to excite the PTT. A pulsed laser source (repetition rate: 60 MHz, pulse width: 3.7 ps) was also employed to compare the level of photo thermal changes with that induced by the CW light. The wavelengths of both the CW sources and the center wavelength of the picosecond optical pulses were around 1550 nm. Polarization controllers (PCs) were employed to ensure the TE polarization of the input light. An optical isolator was inserted in the high-power light path to prevent the reflected light from damaging the light source. The input laser was split into two beams via a 50:50 beam splitter, with one injecting into the device under test (DUT) and the other being sent to an optical power meter (OPM) for monitoring the input power. Lensed fibers were used to butt couple light into and out of the DUT with a coupling loss of ~2 dB / facet. A charged-coupled device (CCD) camera was placed above the waveguide to monitor the adjusting of the coupling. Another OPM was employed to measure the output power after passing through the DUT.

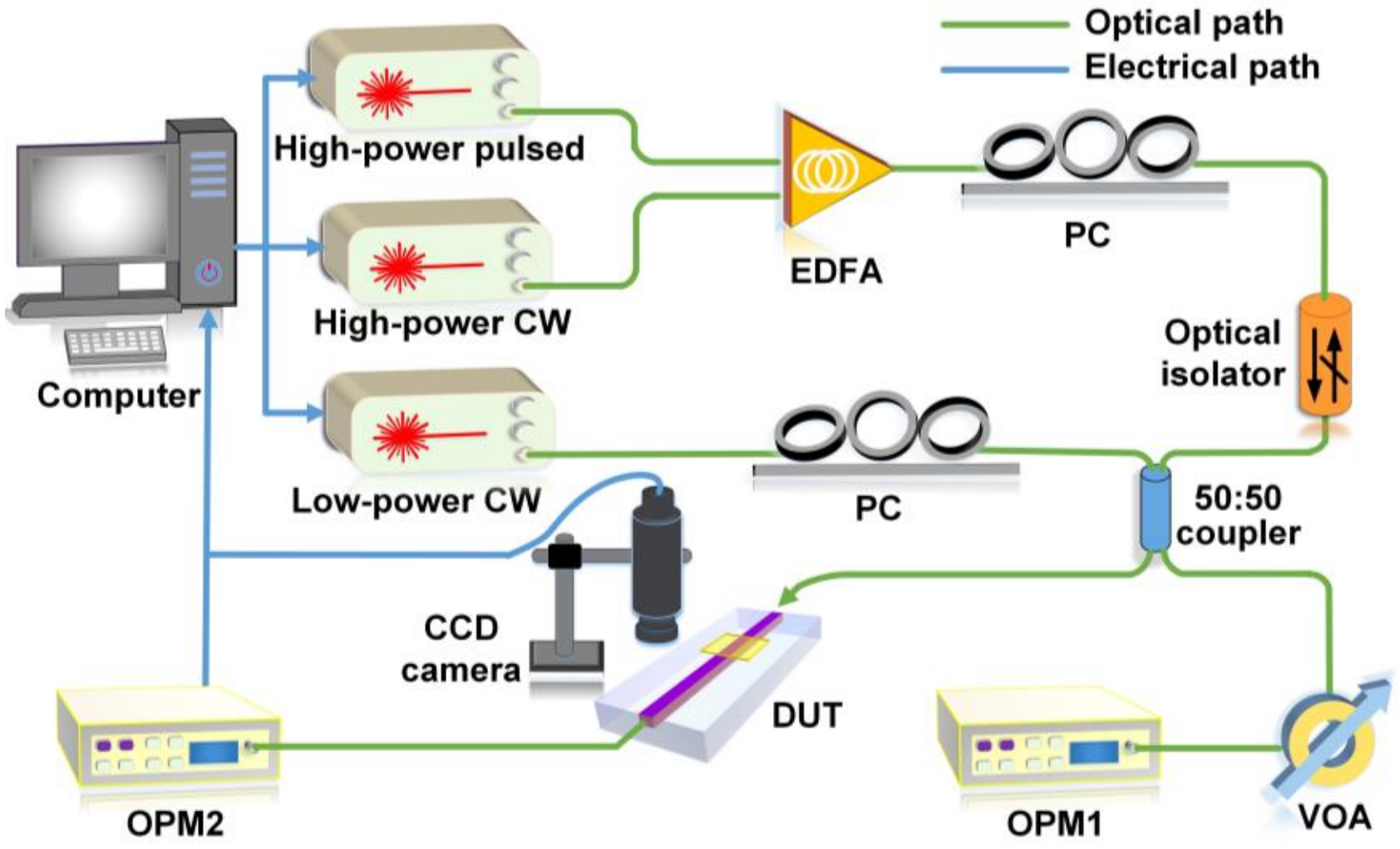

**Figure 2.** Experimental setup for characterization of PTT of GO-coated integrated waveguides. EDFA: erbium-doped fiber amplifier. PC: polarization controller. DUT: device under test. CCD: charged-coupled device. VOA: variable optical attenuator. OPM: optical power meter.

**Figure 3a** depicts the measured insertion loss of the integrated waveguide coated with 1 layer of GO versus input CW power. Unless otherwise specified, the input power of CW light or optical pulses in this paper represents the power coupled into the waveguide after subtracting the fiber-to-chip coupling loss. In order to characterize both the reversible and permanent changes of the material properties, after each measurement at a specific input power, we turned off the high-power CW light and remeasured the insertion loss using a low-power CW light with a power of 0 dBm. The results measured using the high-power and the low-power CW light sources are shown by the red and blue dots, respectively. As can be seen, the evolution of the PTT of the GO film can be divided into three reduction stages. At Stage I, when the input power was below 20 dBm, the insertion loss of the waveguide remained constant despite the increase in input power, reflecting that there was negligible change in the absorption of the GO film and the light power was not high enough to induce obvious photo-thermal changes. At Stage II starting from 20 dBm, the insertion loss showed a slight but observable increase with the input power, indicating the occurrence of the photo-thermal changes in the GO film. The results measured using low-power CW light after tuning off the high-power CW light remained constant. This reflects the fact that there were no permanent changes in the GO films, and the photo-thermal changes at this stage were reversible. These features of the photo-thermal changes in the GO films are consistent with previous reports [19, 34]. For Stage III, when the input power was further increased to above 22 dBm, the results measured using low-power CW light also showed an obvious increase with input power. Since permanently reduced GO films did not show any obvious power dependence [7, 8], this reflects the fact that there were permanent changes in the GO films. In addition, the difference between the red dots and their corresponding blue dots indicates that there was still reversibly reduced GO and only part of the GO film was permanently reduced. We infer that there would be a new stage after Stage III at even higher powers, where the difference between the red and blue dots at the same power would vanish due to the full reduction of all the GO films. We could not observe this stage in our experiments since we had already applied the maximum experimentally available power to the DUT.

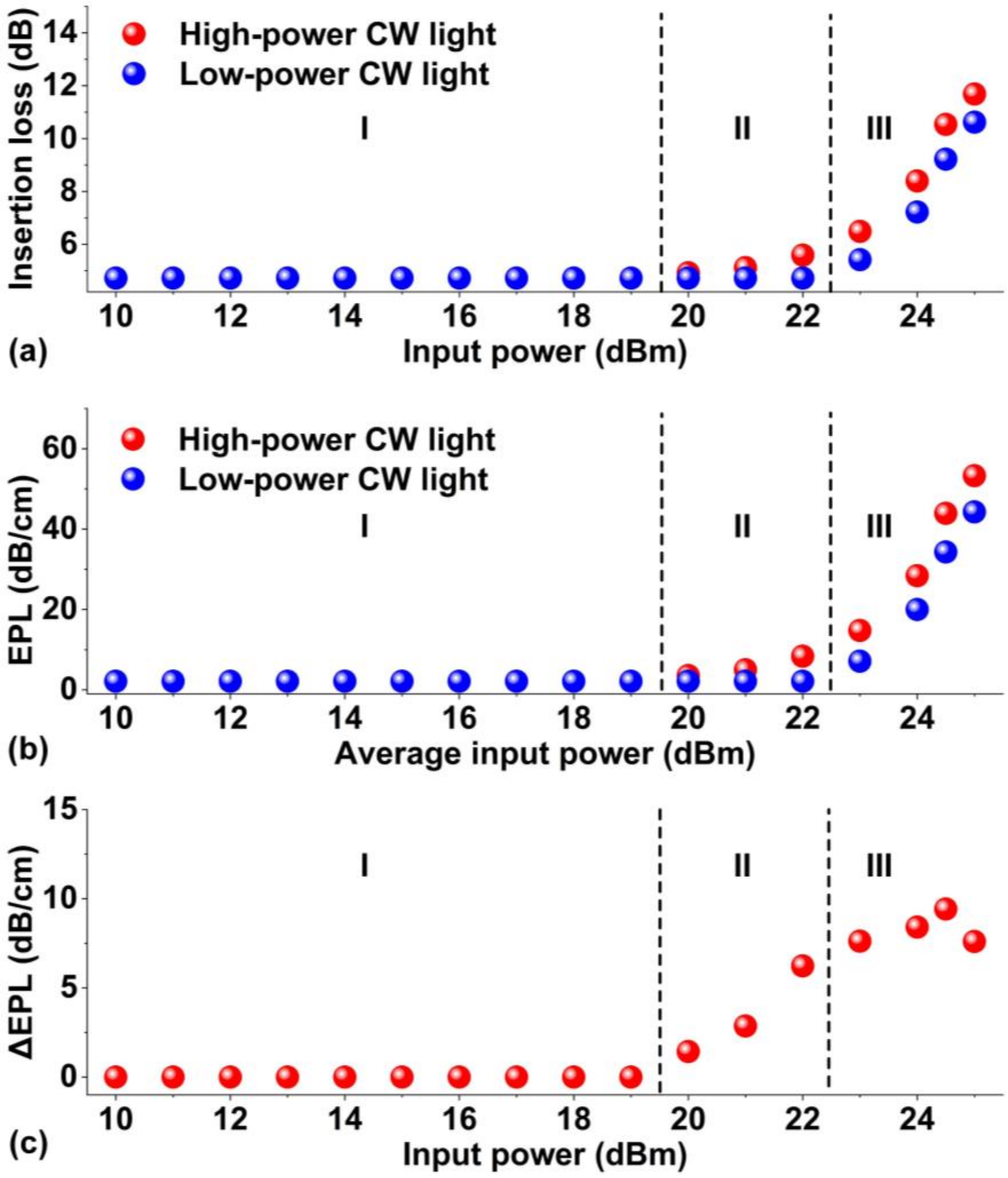

**Figure 3.** Experimental results for characterizing PTT of the hybrid waveguide coated with 1 layer of GO. (a) Insertion loss of the hybrid waveguide versus the input power of the high-power light source. The red dots show the loss of the high-power CW light source, and the blue dots show the loss measured with a low-power CW light source after exposure at the power level indicated on the x-axis. (b) GO-induced excess propagation loss (EPL) versus input power of the high-power light source. The red and blue dots show the results corresponding to the red and blue dots in (a), respectively. (c) $\Delta EPL$ extracted from (b) showing the difference between the red and blue dots.

**Figure 3b** depicts the GO-induced excess propagation loss (EPL) extracted from **Figure 3a**. The *EPL* (dB/cm) is defined as

$$EPL = (IL - IL_0)/L_w \tag{1}$$

where *IL* is the measured insertion loss of the hybrid waveguide in **Figure 3a**, $IL_0$ is the insertion loss of the bare waveguide, and $L_w$ is the GO film length. **Figure 3c** shows the $\Delta EPL$ extracted from **Figure 3b**, which is defined as the difference between the red and blue dots after exposure by the high-power CW source at the input power indicated on the x-axis. As can be seen, the $\Delta EPL$ remained zero at Stage I, and started to increase at Stage II. In Stage III, the $\Delta EPL$ first slightly increased and then decreased when the input power was above 24 dBm. This can be attributed to the hybrid nature of GO films at this stage due to the co-existence of the permanently and reversibly reduced GO.

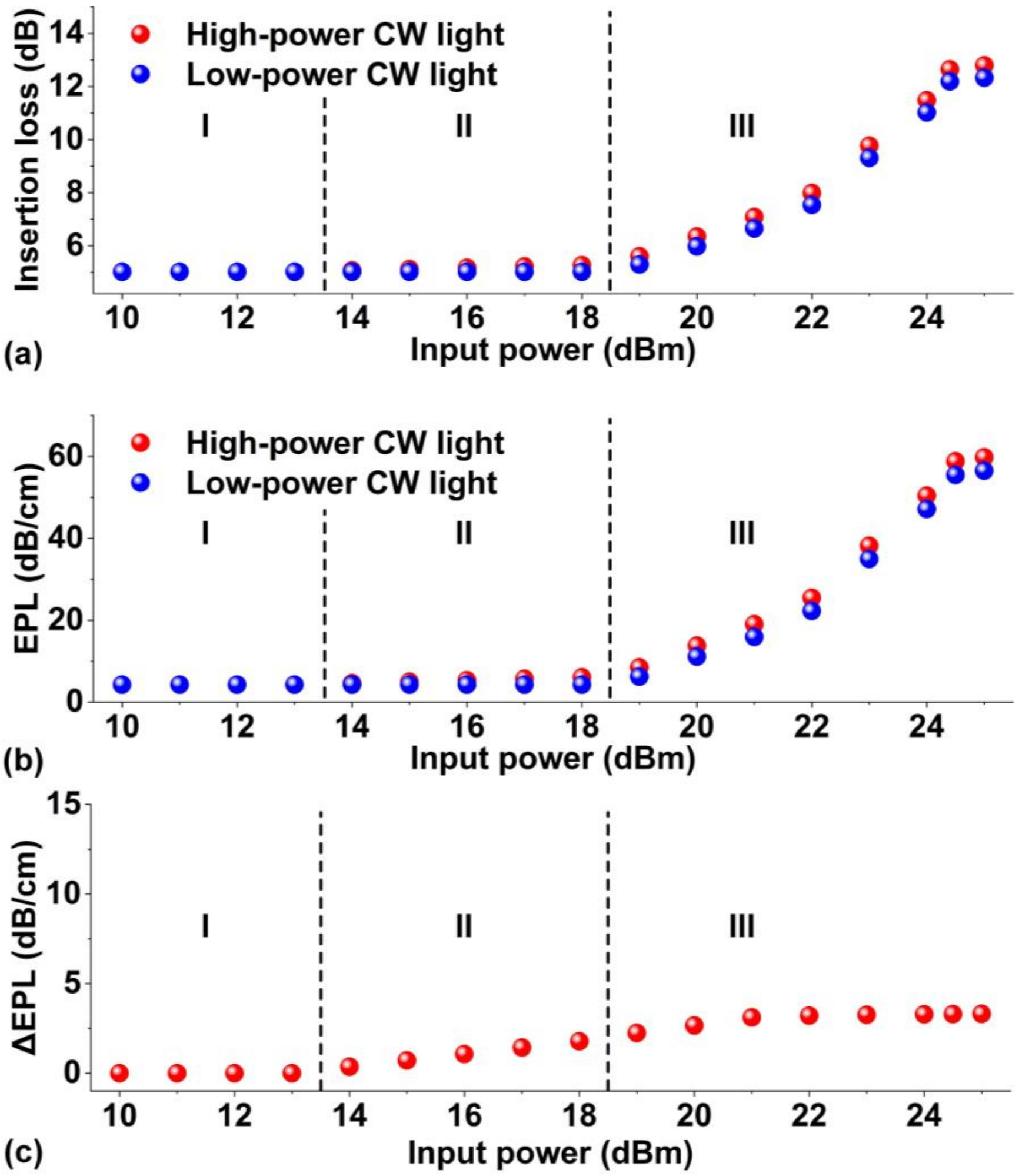

**Figure 4.** Experimental results for characterizing PTT of the hybrid waveguide coated with 2 layers of GO. (a) Insertion loss of the hybrid waveguide versus the input power of the high-power light source. The red dots show the loss of the high-power CW light source, and the blue dots show the loss measured with a low-power CW light source after exposure at the power level indicated on the x-axis. (b) GO-induced excess propagation loss (EPL) versus input power of the high-power light source. The red and blue dots show the results corresponding to the red and blue dots in (a), respectively. (c) $\Delta EPL$ extracted from (b) showing the difference between the red and blue dots.

The corresponding experimental results for the hybrid waveguide coated with 2 layers of GO are shown in **Figure 4**. **Figure 4a** shows the measured insertion loss versus input CW power. Similar to the results for the device with 1 layer of GO, the evolution of the PTT of the device with 2 layers of GO can also be divided into three reduction stages with increasing input power. Compared to the results in **Figure 3a**, the power thresholds for Stage II and Stage III were lower, with Stage II starting at 13 dBm and Stage III starting at 18 dBm. This reflects the fact that the power endurance of the film with 2 layers of GO was lower than the film with 1 layer of GO.

**Figure 4b** shows the GO induced EPL extracted from **Figure 4a**. Compared to the results in **Figure 3b**, the EPL here increased more slowly with input power. At Stage I, the excess propagation loss induced in the film with 2 layers of GO is about twice that induced in the film with a single layer of GO. However, at Stage II and Stage III, the difference between the EPL induced in 1 and 2 layers of GO became smaller, particularly at higher powers above 20 dBm. This is because the EPL defined in **Equation (1)** is a parameter averaged over the GO film length, whereas the reduction of GO in practical hybrid waveguides induced by the photo thermal changes is nonuniform, *i.e.*, the GO film at the beginning of the waveguide is more easily to be reduced, which absorbs more light power and so protects the GO film following it from being reduced. The film with 2 layers of GO absorbed more light than the film with 1 layer of GO, and so the light transmission was more attenuated over a shorter distance, resulting in a higher proportion of unreduced GO. The loss of the non-reduced 2 layers of GO was lower than the single layer of rGO, thus resulting in a lower EPL.

**Figure 4c** shows the $\Delta EPL$ extracted from **Figure 4b**. Unlike the trend in **Figure 3c**, the $\Delta EPL$ here increases with input power in Stages II and III without showing obvious decrease. This is because for the hybrid waveguide with 2 layers of GO, only a small length of the GO film near the waveguide input was permanently reduced, leaving significant lengths of GO films that were either non-reduced or only reversibly reduced. Although the $\Delta EPL$ increased to above zero at relatively low input power for the device with 2 layers of GO, it increased more slowly with input power, with the values at high input powers being smaller than those for the device with 1 layer of GO. This is because for the film with 2 layers of GO, the reversibly reduced GO experienced relatively lower power (14 – 18 dBm), while the film exposed to higher power was permanently reduced and no longer exhibited $\Delta EPL$. On the other hand, for the single layer GO film, the reversibly reduced GO experienced relatively higher powers (20 – 22 dBm), thus yielding a higher $\Delta EPL$.

## 4. Discussion

In section 3, the inhomogeneous GO reduction along the hybrid waveguides was used to explain several experimental phenomena. To verify the inhomogeneous nature of the GO reduction induced by the photo thermal changes, we characterized the Raman spectra of 2 layers of GO coated on an integrated waveguide after applying the maximum input power of 24 dBm. The results are shown in **Figure 5**, where representative D and G peaks of GO can be clearly identified. Near the waveguide input, the D and G peaks of the detected Raman signals are relatively small, with a D/G ratio being larger than 1. This is similar to that of graphene [11], indicating that the GO film here was deeply reduced. In contrast, for positions further away from the waveguide input, the D/G ratio decreased to become less than 1, together with an increased intensity for the detected Raman signal. These characteristics show agreement with Raman spectra of GO having fewer defects [9, 12] and reflect the fact that the GO away from the waveguide input was reduced less.

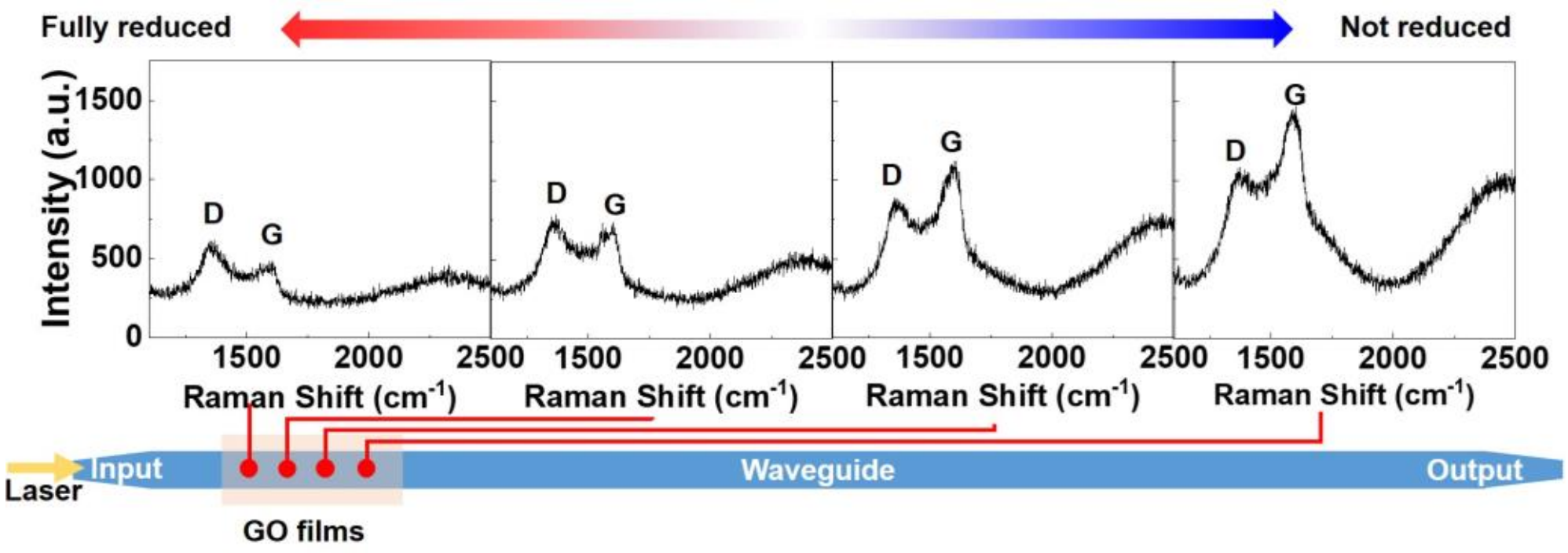

**Figure 5.** Raman spectra of 2 layers of GO coated on an integrated waveguide after applying an input CW power of 24 dBm.

To compare the PTT induced by CW light versus optical pulses, we measured the EPLs of the hybrid waveguides with a single GO layer for both CW light and optical pulses having the same average power. The results are shown in **Figure 6**. The optical pulses had a repetition rate of ~60 MHz and a pulse width of ~3.7 ps, which corresponded to a peak power $4 \times 10^3$ times higher than the CW light with the same average power. As can be seen, both the CW light and the optical pulses induced measurable EPLs at high average input powers. The small difference between them indicates that the EPL was mainly a function of the average power rather than peak power. This is in agreement with observations in GO films arising from photo-thermal processes in Refs. [19, 29, 30], and further confirms the existence of the photo thermal changes. In contrast, the changes induced by ultrafast nonlinear optical processes such as four-wave mixing, two-photon absorption, and saturable absorption are dependent on the peak input light power [34, 39, 40]. The slightly lower EPL induced by optical pulses compared to CW light can be attributed to saturable absorption in the GO films caused by the high peak powers, which was also observed in Refs. [23, 41]. We also measured permanent EPLs with low-power CW light (0 dBm) after turning off the high-power CW light and optical pulses. The permanent EPLs induced by the CW and optical pulses showed negligible difference, reflecting the fact that the permanent reduction of GO was mainly induced by the photo thermal changes. These results have significant implications for designing and engineering nonlinear photonic chips that would benefit from the integration of GO films for applications such as classical [42-90] and quantum optical microcombs. [90-99]

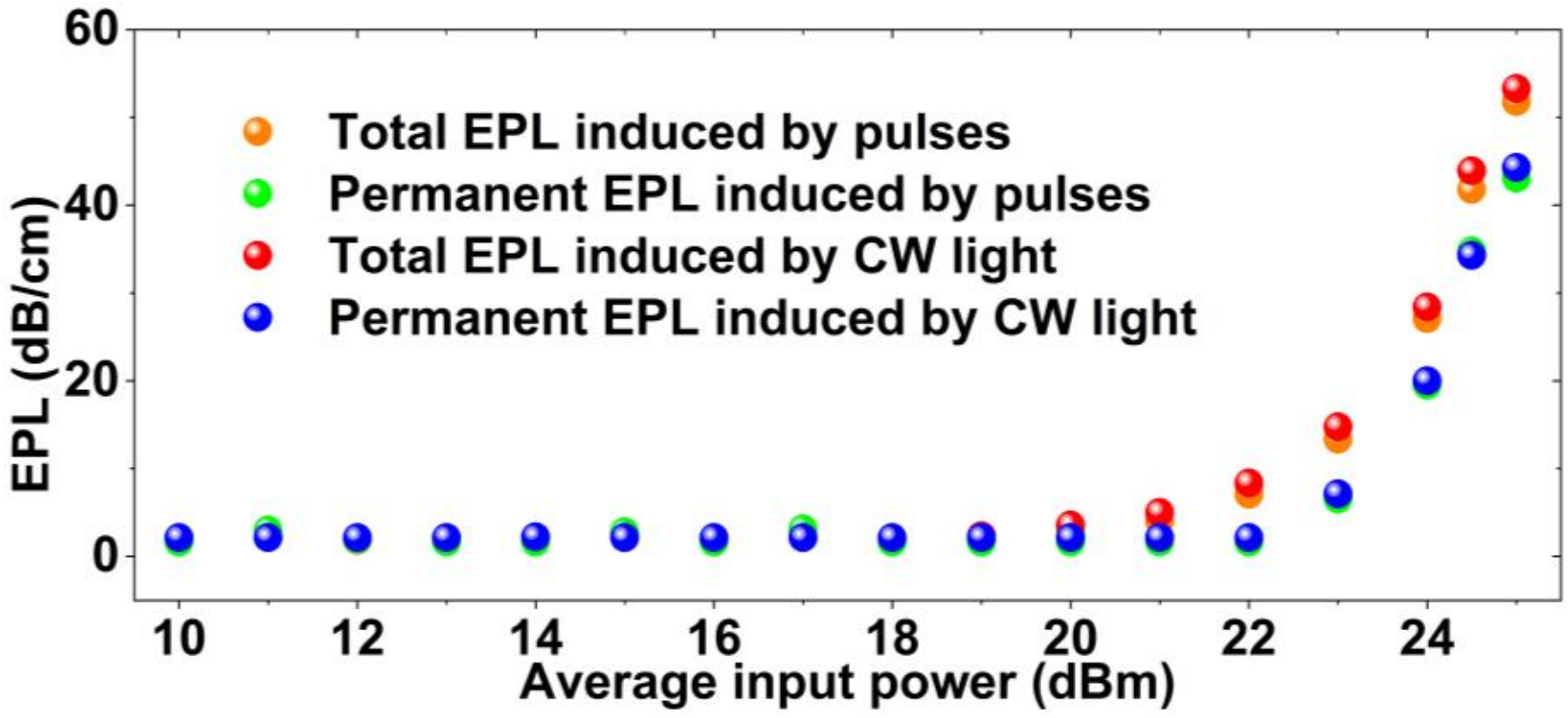

**Figure 6.** Experimental results of the total and permanent EPL induced by a CW light and optical pulses versus average input power for the hybrid waveguides coated with 1 layer of GO.

## 5. Conclusions

In this paper, we present detailed characterization of the PTT of GO films coated on integrated optical waveguides. Reversible and permanent GO reduction are observed by applying different CW laser powers to the devices with 1 and 2 layers of GO. The corresponding power thresholds are measured, with three reduction stages being identified. The Raman spectra at different positions of a hybrid waveguide with a permanently reduced GO film are characterized, which verifies the inhomogeneity of GO reduction. The photo-thermal changes induced by CW light and optical pulses with the same average power are also compared, which show negligible difference and confirms that the PTT mainly depends on the average input power. These results are useful for controlling and engineering the material properties of GO in hybrid integrated photonic devices.

**Conflicts of Interest:** The authors declare no conflict of interest.